**New Physics Searches Using Precision Spectroscopy**

**Author:** Chad Orzel

**Affiliation:** Union College Department of Physics and Astronomy, Science and Engineering Center, Schenectady, NY 12308 USA

**Abstract:** The exceptional precision attainable using modern spectroscopic techniques provides a promising avenue to search for signatures of physics beyond the Standard Model in tiny shifts of the energy levels of atoms and molecules. We briefly review three categories of new-physics searches based in precision measurements: tests of QED using measurements of the anomalous magnetic moment of the electron and the value of the fine-structure constant, searches for time variation of the fundamental constants, and searches for a permanent electric dipole moment of an electron or atomic nucleus.

## 1) Introduction

The Standard Model of particle physics occupies a unique position in the history of science, as arguably the most successful theory that is nevertheless known to be incorrect. The Standard Model framework represents the culmination of decades of work, tying together electromagnetic, weak nuclear and strong nuclear interactions and explaining the structure of matter through its six quarks, six leptons, and associated force-carrying bosons. Predictions within the Standard Model framework have been experimentally verified across an enormous range of energy scales, and confirmed at the parts-per-trillion level (Gabrielse, 2013).

On the other hand, though, we have ample evidence that the Standard Model is incomplete. A wide range of observations from astronomy and cosmology suggest that the ordinary matter described by the Standard Model is only around 4% of the energy content of the observable universe, with about 20% of the total consisting of non-baryonic "dark matter" (Bertone et al., 2005) and 76% of the energy in the form of "dark energy" (Frieman et al., 2008). Neither dark sector component is adequately described by the Standard Model.

The observation that the visible universe does not contain large amounts of antimatter suggests that the Big Bang created substantially more matter than antimatter, which requires violation of charge and parity (CP) symmetry (Sakharov, 1967). While the Standard Model framework does include sources of CP-violation, they are not enough to explain the observed abundance of matter over antimatter.

Finally, a theory of quantum gravity that successfully merges the Standard Model particles and forces with the curved spacetime of General Relativity remains elusive. Numerous theoretical approaches have been tried over a few decades (Kiefer, 2005; Elvang and Horowitz, 2014; Ashtekar, et al., 2014), but despite considerable effort, there is no consensus as to how best to quantize gravity.

All of these difficulties are interrelated, and suggest that the universe must contain particles and fields beyond those known in the Standard Model. Attempts to explain dark energy as a manifestation of the quantum vacuum energy of the Standard Model fields fail by many orders of magnitude, suggesting that the explanation may require a theory of quantum gravity, and attempts to unify gravity with other forces almost inevitably introduce additional fields to the theory. These new fields generally introduce additional source of CP-violation that can help explain baryogenesis, and a long-lived but as yet undetected massive particle from beyond the Standard Model would also offer an attractively simple explanation for the observation of dark matter.

These observations, among other factors, have led to a large and active experimental effort to find evidence of physics beyond the Standard Model. The best-known of these are collider experiments attempting to directly create new particles, most notably the Large Hadron Collider. Despite some intriguing hints (see, for example, ATLAS, 2017), these have yet to provide clear evidence of new physics.

In parallel with attempts to actively create new particles in high-energy colliders, numerous groups are searching for new physics through passive observations. If dark matter takes the form of long-lived particles from beyond the Standard Model, and these particles interact with ordinary matter through forces other than gravity, it should be possible to detect these interactions. Numerous experiments to directly detect particle dark matter are underway or in preparation; as yet these have not produced a definitive detection (Liu et al., 2017).

The subject of this review is a third approach to the search for physics beyond the Standard Model, using measurements of atomic and molecular properties to indirectly detect the presence of new particles and fields. The idea of spectroscopic measurements leading to new physics has deep historical roots: the discovery of the Lamb shift (Lamb and Retherford, 1947) and the anomalous magnetic moment of the electron (Nafe et al., 1947; Nafe and Nelson, 1948) in the late 1940's spurred the development of quantum electrodynamics (QED), which is one of the cornerstones of the modern Standard Model.

In a modern context, using atoms and molecules to detect new particles may initially seem like an improbable tactic, given the vast gulf between the energy scales involved. The new particles sought at the LHC and in most direct-detection searches for dark matter have masses of order 0.1-1 TeV, and the contribution of TeV scale physics to intra-atomic or –molecular energies (typically of order 0.1-1eV) should be extremely small. As wide as this gulf is, though, the measurement precision attainable with modern spectroscopic techniques-- approaching a part in $10^{18}$ for state-of-the-art atomic clocks-- makes the detection of even the minuscule energy shifts expected from new physics feasible and opens the possibility of probing physics beyond the Standard Model in tabletop experiments.

Interest in the use of precision measurements to search for new physics has exploded over the last two decades, and grown into a thriving subfield of atomic, molecular, and optical physics. Several recent articles have been published offering an overview of all or part of the field; the most comprehensive of these is Safronova et al., 2017. In this article, like others (DeMille et al., 2017; Karshenboim and Ivanov, 2017), we will focus on selected parts of the field, looking at three particular areas: measurements of the anomalous magnetic moment of the electron and thus the fine-structure constant α that provide a direct test of QED; comparisons between atomic clocks that search for time variation of the fundamental constants; and searches for forbidden moments of particles and nuclei that provide a measurement of time-reversal violation (and thus CP-violation).

All of these searches ultimately derive their power from techniques developed for use in atomic spectroscopy and atomic clocks. Thus, we will begin with a brief discussion of the general problem of spectroscopy and how it connects to fundamental physics. Next, we will briefly review Ramsey interferometry, femtosecond frequency combs, and their applications in microwave and optical atomic clocks and comparisons between them. Then we will review specific experimental programs, and conclude with some discussion of future prospects for the field.

## 2) Spectroscopy and Fundamental Physics

The central concern of precision spectroscopy is to determine the frequency of light needed to drive a quantum system—an atom, a molecule, or a trapped particle—between two energy eigenstates, and to determine how those transition frequencies change in response to external applied fields. To achieve high precision in measurements of the transition frequency, these must be states with low rates of spontaneous emission, and thus low natural linewidths. Typical experimental states for precision measurements are the hyperfine ground states of atoms, or dipole-forbidden transitions between metastable electronic states, that proceed through higher-order processes.

As these energy eigenstates are primarily determined by the electromagnetic interaction between the constituent particles, it may not be immediately obvious how physics beyond the Standard Model can enter into these systems. After all, the electromagnetic interaction is a central component of the Standard Model, and has been well understood for decades.

The full solution for even a hydrogen atom, though, requires a relativistic quantum field theory—even the relativistic Dirac solution for the hydrogen energy levels does not account for the Lamb shift between the $^2S_{1/2}$ and $^2P_{1/2}$ levels. In the QED framework, the energies for bound electronic states are approximated as a series in increasing powers of the fine structure constant

$$\alpha = \frac{e^2}{4\pi\epsilon_0 \hbar c} \approx \frac{1}{137} \tag{1}$$

This offers two paths by which new physics can enter the determination of energy levels in atoms and molecules. First, in some extensions of the Standard Model, the coupling constant α itself becomes a dynamical variable, and can change over time. Such a variation would lead to a shift in energy levels over time, which can in principle be detected either retrospectively through comparisons between modern and historical measurements, or through repeated laboratory measurements to constrain present-day variation. (These will be discussed in Section 5).

Second, the new particles and fields introduced by extensions to the Standard Model will appear as virtual particles in higher order Feynman diagrams in the QED expansion. These can be detected by either a difference between the energies predicted using Standard-Model QED and the experimentally measured values, or through the appearance of symmetry-violating interactions (which will be discussed at greater length in Section 6).

In the specific case of atomic hyperfine transitions, the interaction between the electronic and nuclear magnetic moments adds additional connections to fundamental physics. The hyperfine splitting can be expressed as:

$$\Delta E_{hfs} = hc\, R_\infty A_{hfs} g_i \left(\frac{m_e}{m_p}\right) \alpha^2 F_{hfs}(\alpha) \tag{2}$$

where $R_\infty$ is the Rydberg constant, $g_i$ the nuclear g-factor, and $A_{hfs}$ and $F_{hfs}$ numerical factors specific to the atom in question. This energy difference depends on the ratio $\mu = m_e/m_p$ between electron and proton masses; with the addition of new particles and fields that couple to the Standard Model leptons and quarks, this mass ratio also becomes a dynamical variable in many theories of physics beyond the Standard Model. The potential for changes in the mass ratio over time introduces another source of possible time variation in atomic energy levels that can in principle be detected spectroscopically.

A number of practical issues must be taken into account when considering atomic and molecular systems as potential systems for new-physics searches. As noted above, the states involved should be long-lived ones, to avoid measurement uncertainties associated with spontaneous emission. The energy states in question must also be addressable with currently available lasers and optics, which cover a range of wavelengths from around 200-2000nm. The sensitivity to any particular new physics channel can also vary enormously between transitions, even within the same atom or molecule; for most beyond-Standard Model physics, the sensitivity tends to increase with atomic number Z, so most experiments involve heavier atoms. This sensitivity must be determined from theoretical calculations connecting the atom-scale observables to the properties of fundamental particles. This tends to limit the systems of interest to few-electron atoms and relatively simple molecules, which are more theoretically tractable.

While this collection of requirements may seem highly restrictive, given the enormous range of elements in the periodic table, there is no shortage of suitable systems for new physics searches. This set of requirements does mean that the field tends to advance in "punctuated equilibria," to borrow a term from biology, with new experimental or theoretical techniques suddenly opening a new category of systems, which is explored with several different experiments before a local optimum is located. Thus, many new-physics searches were originally carried out in neutral atoms before moving to polar molecules or ions once more sophisticated techniques to identify and manipulate the relevant systems were developed.

**3) Frequency Measurements and Atomic Clocks**

The late Arthur Schawlow famously advised his students to "never measure anything but frequency" (Hänsch, 2006), and indeed, the potential for detecting new physics using atoms and molecules is rooted in the impressive precision that can be attained when measuring frequency. This is reflected in the SI system of units: the second is defined in terms of a frequency, as 9,192,631,770 oscillations of the light associated with the ground state hyperfine splitting in cesium. Since 1983, the meter has been defined as $1/299,792,458^{th}$ of the distance traveled by light in one second, converting the standard of distance into a measurement of time, and thus frequency. And after many years of development (Richard et al., 2015), in 2018 the BIPM is expected to formally redefine the kilogram in terms of Planck's constant, implicitly connecting the rest energy of a particle to a frequency. When that redefinition is complete, time, distance, and mass will all be based on measurements of frequency.

Modern techniques for frequency measurements and comparisons were developed in the context of atomic clocks. In this section, we will briefly review the essential frequency metrology techniques of Ramsey interferometry for frequency measurement and femtosecond frequency combs for comparing frequencies across wide ranges of the spectrum. Then we will review the principal technologies used in modern frequency standards, including the fountain clocks that are used as current primary standards, and the trapped ion and optical lattice clocks that are being developed for a possible future generation of time standards.

3.1) Ramsey Interferometry

The separated-field spectroscopy technique was developed by Norman Ramsey in 1950 (Ramsey, 1950), in the context of atomic or molecular beams interacting with microwaves. Rather than a single long-duration interaction between light and the reference atoms, for which it becomes technically challenging to maintain uniformity of the fields over a long flight path, it uses two briefer interactions, and an interference between different parts of an evolving superposition. The essential technique is quite general, and variations on it are at the heart of most high-precision spectroscopic measurements across a wide range of frequency, from the microwave to ultraviolet regions of the spectrum.

In Ramsey interferometry, a system in state |1> is first exposed to a pulse of light at frequency $\omega$, near the resonant frequency $\omega_0$ that will drive the system to state |2>. The system will undergo coherent Rabi oscillations at a frequency $\Omega$, and the light intensity and pulse duration are chosen to give a "$\pi/2$-pulse," leaving the system in an equal superposition of states |1> and |2>. This superposition is allowed to evolve freely for a time T, then a second $\pi/2$-pulse is applied. At the end of this pulse sequence, the state of the system is probed to determine whether it has made a transition from state |1> to state |2>.

Integration of the Schrodinger equation for this pulse sequence shows that the probability of a transition after the second pulse is:

$$P(1 \to 2) = \left(\frac{\Omega \tau_{\pi/2}}{2}\right)^2 \left[\frac{\sin\left(\frac{\omega-\omega_0}{2}\right)\tau_{\pi/2}}{\left(\frac{\omega-\omega_0}{2}\right)\tau_{\pi/2}}\right]^2 \cos^2\left(\frac{\omega-\omega_0}{2}T\right) \tag{3}$$

This consists of an overall amplitude depending on the Rabi frequency $\Omega$ and the pulse duration $\tau_{\pi/2}$ modulated by an oscillating factor depending on the free evolution time T and the frequency of the applied light. The frequency width of the "Ramsey fringes" resulting from this term is $\Delta\omega=\pi/T$, inversely proportional to the free evolution time T, and is the key to the extreme sensitivity of the method.

The separated-fields method is at its core an interferometric technique, relying on the evolving phase difference between the two states in the superposition. This is commonly illustrated using the Bloch sphere picture, shown in Fig. 1, making an analogy between the two-level system of interest and the magnetic moment of a spin-1/2 particle. The state of the system is represented as a vector to a point on the surface of a unit sphere, where the south and north poles correspond to states |1> and |2>, respectively, and the polar angle of the vector describes the admixture of the two states in a superposition. States on the equator of the Bloch sphere represent an equal superposition of |1> and |2>, and the azimuthal angle describes the relative phase between the two components.

The initial $\pi/2$-pulse of the Ramsey sequence rotates the state vector about the x-axis, bringing it from the south pole to the equator, along the y-axis. During the free evolution time, the phase of this superposition will evolve at a frequency $(\omega - \omega_0)$ (in the rotating wave approximation), which corresponds to a precession of the state vector about the z axis. The second $\pi/2$-pulse performs another rotation of the state vector about the x-axis, in the same direction as the first.

If the precession during the free evolution time T amounts to an integer number of revolutions, so that the state vector is again pointing along the y-axis, the second $\pi/2$-pulse completes the transition from |1> to |2>, and the transition probability is P(1→2)=1. If the precession amounts to a half-integer number of revolutions, though, the state vector is along the –y-axis, and the second $\pi/2$-pulse returns it to the initial state (P(1→2)=0).

The narrow Ramsey fringes, then, are a result of the precession of the state vector driven by the evolving phase difference between the two terms of the superposition. In a frequency standard or clock, the oscillating probability for a fixed T determines the detuning of the applied field, and a correction fed back to the source creates a stable frequency referenced to the atomic energy levels of interest. This process offers the most precise method known for determining the frequency associated with energy differences between quantum states, and most precision AMO experiments make use of some version of Ramsey interferometry.

3.2) Femtosecond Frequency Combs

As the SI second is defined in terms of the hyperfine splitting of the $^{133}$Cs ground state, any measurement of an absolute frequency necessarily involves a comparison to a cesium standard. For atomic transitions in the optical domain, though, a direct comparison to the microwave frequency in cesium presents a significant technical challenge. In recent years, this process has been greatly simplified with the development of "frequency comb" sources based on femtosecond pulsed lasers.

A short-duration pulse requires the addition of Fourier components spanning a wide frequency bandwidth; for a sufficiently short laser pulse (a few femtoseconds at optical frequencies), the bandwidth can span a full octave. Frequency comb sources have found applications in molecular spectroscopy and as reference sources for astronomical spectrometers, but our primary interest in them here is as a tool enabling high-precision comparisons of laser frequencies.

The frequency spectrum of a mode-locked laser will contain a large number of regularly spaced allowed modes with the spacing determined by the length of the laser cavity (which also determines the repetition rate of the laser). The frequency of the nth mode in the comb is:

$$\nu_n = n\, \nu_{rep} + f_{cav} \quad (4)$$

where $f_{cav}$ is a frequency offset due to dispersion within the cavity, and $\nu_{rep}$ is the repetition rate of the laser, typically of order 100MHz. An octave-spanning comb will have modes whose frequencies differ by a factor of two; light from the lower of the two can be frequency doubled and mixed with light from the higher-frequency mode producing a beat note at the difference frequency (Fig. 2):

$$\Delta\nu = 2\nu_n - \nu_{2n} = (2n\nu_{rep} + 2f_{cav}) - (2n\, \nu_{rep} + f_{cav}) = f_{cav} \quad (5)$$

This beat note directly measures the cavity offset, and thus allows the determination of the absolute frequency of any mode in the comb.

A self-referenced frequency comb greatly simplifies the process of comparing transition frequencies in the optical range to microwave frequency standards. The repetition rate, cavity offset, and the beat note between a laser locked to the transition of interest and the nearest comb mode will all be in the RF range and thus readily compared to microwaves derived from atomic clocks. The combination of the three gives the absolute frequency of the atomic transition.

The need to compare optical to microwave sources necessarily means that any absolute frequency determination is limited to the same precision as the atomic clock, around a part in $10^{16}$ for current cesium standards. (This could change if the SI second were to be redefined in terms of a different atomic transition, but this is not expected to happen in the near future.) The inherent uncertainty of optical-frequency standards can be better than this, though, thanks to the higher transition frequency, and octave-spanning combs allow frequency *comparisons* between different optical standards at the level of a few parts in $10^{19}$. For such comparisons, the comb is stabilized with reference to one of the two optical transitions, and the beat note between the other laser and the nearest comb mode is measured to determine the ratio of laser frequencies to some 18 decimal places.

**3.3) Fountain Clocks**

The original realizations of the Ramsey separated-fields scheme used an atomic beam passing through two separate microwave cavities. This method is limited in its precision by the difficulty of providing a long free evolution time in a beam of atoms moving at thermal velocities, and also by the technical challenge of fabricating two identical microwave cavities.

The best current realizations of Cs-based frequency standards use a "fountain" geometry, with an atomic sample passing through the same physical cavity twice. This was originally proposed by Zacharias in the 1950's but only became practical with the development of laser cooling techniques in the 1980's. A cloud of Cs atoms in the F=3 ground state are launched upward, making an initial pass through the cavity for the first π/2 pulse of the Ramsey sequence. The free evolution period occurs as the atom cloud rises to its maximum height decelerating under the influence of gravity. The atoms then fall back down through the microwave cavity a second time, which completes the Ramsey pulse sequence.

The free evolution time T for a fountain clock is determined by the maximum height of the launched atoms. For typical modern clocks, this is around 1m, so the total evolution time is about T=1s. Ramsey interferometry allows frequency resolution of around 1Hz on the measurement of the 9.19 GHz Cs hyperfine splitting. The overall clock performance improves with some averaging, eventually reaching a statistical uncertainty of a few parts in $10^{-16}$ (Heavner et al., 2014, Levi et al., 2014).

This represents the culmination of many decades of work that have improved the Cs clock uncertainty by almost three orders of magnitude since 1980. Further progress is likely to be difficult, though, as at the $10^{-16}$ level the uncertainty is dominated by tiny and hard to characterize systematic effects. For example, recent literature on atomic clocks has included discussion of a possible "microwave lensing" shift caused by deflection of the atomic wavepackets during their interaction with the light field (Jefferts et al., 2015).

**3.4) Optical Clocks**

Any dramatic future improvement in the development of time standards is likely to involve a shift away from microwave standards based on hyperfine transitions to optical standards based on dipole-forbidden electronic transitions. For a comparable measurement uncertainty $\Delta \nu$, the fractional uncertainty $\Delta \nu / \nu$ will necessarily be smaller with optical clocks, as the transition frequencies are some five orders of magnitude larger than for microwave standards.

Numerous candidate systems for an optical-frequency time standard are under investigation in labs around the world (Ludlow et al., 2015). Optical clocks at the frontier of precision measurements can be categorized into two general approaches: one class using trapped ions, the other neutral atoms in optical lattices.

Trapped ion standards use small numbers of ions, often single ions, tightly confined in a Paul trap made from rapidly switched high-voltage potentials. Ions are laser cooled to the motional ground state of the trapping potential, where the tight confinement prevents frequency shifts associated with motion of the ion after absorbing and emitting photons. Changes of internal state are detected using fluorescence on the laser cooling transition, as one of the clock states readily absorbs cooling light while the other does not. The π/2 pulses for Ramsey spectroscopy are supplied by additional lasers, usually via two-photon transitions for symmetry reasons, as the clock transitions are generally dipole forbidden.

Numerous ion species have been investigated for clocks, including (but not limited to) Al+ (Rosenband, et al., 2008), Ca+ (Chwalla, et al., 2009; Huang et al., 2012, Matsubara et al., 2012), Sr+ (Dube et al, 2014), Hg+, and Yb+ (Huntemann et al., 2012; King et al., 2012; Godun et al., 2014; Huntemann et al., 2014). Of particular note is the Al+ system, where the clock ion cannot be directly cooled with currently available lasers. Instead, the Al+ system is a "quantum logic clock," making use of a second ion of either Be+ or Mg+ held in the same trap. The Al+ "clock" ion is cooled sympathetically by the "logic" ion, and the state preparation and detection are accomplished with Raman pulses that map the state of the logic ion onto the clock ion (and vice versa) via the common motional state of the two trapped ions.

The Yb+ system is also of special interest, as it features two accessible clock transitions, one an electric quadrupole (E2) transition at a wavelength of 436nm, the other an octupole (E3) transition at 467nm. Both of these have been characterized and measured in a single ion, and will be discussed further below.

Trapped ion systems offer the benefit of long confinement times, allowing a given ion to be interrogated many times, and tight confinement that removes most systematic sources of error associated with the motion of the atom from the measurement of the internal electronic states. The small number of particles in these traps and the strong interactions between ions, however, limits the signal-to-noise achievable for state detection. The other major class of optical clock candidates use neutral atoms, which can be confined in large numbers in optical lattices.

An optical lattice confines atoms to wavelength-scale sites using the light shift (AC Stark shift) of a spatially varying pattern of intensity. The light shift scales as the ratio of laser intensity to detuning ($I/\Delta$), while the light scattering rate scales as $I/\Delta^2$, so with a laser sufficiently far from the atomic resonance, an essentially conservative trapping potential can be produced, confining the atoms without photon scattering. Atoms in these systems can be cooled to the motional ground state within a given lattice site; as with trapped ions, this separates the center-of-mass motion of the atom from measurements of its internal states, greatly reducing many systematic uncertainties associated with free particles..

The use of a trapping potential based on light shifts of the atomic ground state, however, may seem antithetical to the idea of precision measurement of the internal energy states. The light shift of a given electronic state depends on the dipole moment induced by the applied laser field, which will in general be different for the ground and excited state of any particular transition. Thus, the presence of the lattice light will tend to shift the transition frequencies of interest in a way that would appear to make precision spectroscopy impossible.

As pointed out by Katori, though (Katori et al,, 2003), there exist particular combinations of wavelength and polarization for which the light shift of the upper state in the clock transition and the trapped ground state are exactly equal. If the lattice is operated at one of these "magic wavelengths," ground-state atoms experience a light shift lattice potential, but the clock transition frequency is unperturbed by the lattice laser. A magic-wavelength lattice system then offers the benefits of tight confinement found in trapped ion systems, along with the enhanced detection efficiency that comes from confining larger numbers of atoms (Ye at al., 2008).

Several neutral atom species have been investigated for use in optical lattice clocks. the most fully developed of these is strontium, with Sr optical lattice clocks in operation in Tokyo, Boulder, and Paris since 2005 (Takamoto et al., 2005; Ludlow et al, 2006; Le Targat et al., 2006). Several laboratories have also developed lattice clocks using ytterbium and mercury. The best of these standards report fractional frequency uncertainties at the level of a few parts in $10^{18}$. Absolute frequency measurements of these transitions are still limited by the uncertainty of the cesium clocks that define the SI second, but comparisons between optical frequency standards can make use of the full precision of these standards in searches for physics beyond the Standard Model, as discussed below.

4) Tests of QED

As noted above, the energy states of atoms and molecules are ultimately determined by electromagnetic interactions and calculated using quantum electrodynamics. As such, the most direct approach to searching for physics beyond the Standard Model is through a test of QED itself, comparing high-precision experimental measurements to theoretical calculations of the same properties. As any QED calculation ultimately leads to an expansion in powers of the fine-structure constant, this comparison is closely connected to measurements of $\alpha$ itself.

4.1) The Anomalous Magnetic Moment of the Electron

While the Lamb shift is the most famous of the experimental results that demonstrated the need for the development of quantum electrodynamics, another critical discovery was a discrepancy of about 3.4MHz between the predicted value for the hyperfine splitting of hydrogen and the value measured by Nafe et al., (1947). This was quickly attributed to a greater-than-predicted value for the magnetic moment of the electron (Kusch and Foley, 1948), usually expressed in terms of an "anomalous magnetic moment" $a_e = \frac{g-2}{2}$ resulting from the g-factor for the electron being greater than the value g=2 predicted by Dirac's relativistic quantum mechanics. One of the signature successes of QED upon its invention in 1947 was Schwinger's analytical result for the lowest-order QED contribution to the anomalous magnetic moment of the electron (Schwinger, 1948, Schwinger, 1949), $a_e = \frac{\alpha}{2\pi}$; this result is famously engraved on Schwinger's tombstone.

Sixty years later, anomalous magnetic moments remain some of the most important tests of QED, and thus of Standard Model physics. The most sensitive current measurement of g uses an "artificial atom" consisting of a single electron confined in a Penning trap. The Penning trap uses a strong magnetic field to drive cyclotron motion, and a quadrupole electric field confining the electron along the magnetic field axis. The energy of the electron is determined by a combination of motion at the cyclotron frequency ($v_c = \left(\frac{e}{2\pi m_e}\right) B_z$) and the interaction of the spin with the axial magnetic field, shown schematically in Fig. 3. The total energy of the nth cyclotron level is:

$$E_{nm_s} = E_n^{(cyc)} + E_{m_s}^{(spin)} \tag{6}$$

$$= \left(n + \frac{1}{2}\right) h\nu_c + \frac{g}{2} h\nu_c m_s - \frac{1}{2} h\delta \left(n + \frac{1}{2} + m_s\right)^2 \quad (7)$$

where the spin projection $m_s = \pm\frac{1}{2}$ and δ is a relativistic correction factor of order $10^{-9}$. Changes in the cyclotron level or the spin state can be driven by weak RF fields applied through the trap electrodes. The same electrodes are used to pick up the axial motion of the electron, which is weakly coupled to the cyclotron motion allowing direct measurement of the state of the electron.

If the electron were a Dirac point particle, two states differing by both a cyclotron excitation and a spin flip (that is, $|n, m_s = +\frac{1}{2}>$ and $|n + 1, m_s = -\frac{1}{2}>$) would be degenerate to within ~hδ. For a real electron with an anomalous magnetic moment, these two levels differ by an "anomaly frequency" in the RF that can be measured to high precision. The best measurement of $a_e$ to date (Hanneke, et al., 2008; Hanneke et al., 2011) is accurate to 0.3 ppt:

$$a_e = 0.001\ 159\ 652\ 180\ 73(28) \quad (8)$$

Calculating a theoretical value of g from QED requires a tenth-order calculation involving the summation of nearly 13,000 Feynman diagrams. The resulting value (Aoyama et al., 2017) is in excellent agreement with experiment, giving a value of

$$a_e^{(th)} = 0.001\ 159\ 652\ 181\ 031\ (15)(15)(720) \quad (9)$$

(The uncertainties in parentheses come from the electroweak and hadronic contributions to the QED calculation, and the measured value of the fine-structure constant, respectively.) The difference between these is $0.3 \times 10^{-12}$, considerably smaller than the theoretical uncertainty, which is dominated by the uncertainty in the measurement of the fine structure constant (Aoyama et al., 2015; Aoyama et al., 2017b).

4.2) Measurement of α

The uncertainty in the calculated value of the anomalous magnetic moment is dominated by the uncertainty in the measured value of α, which suggests an alternative approach to the interpreting the measurements of Hanneke et al. (Hanneke et al., 2008; Hanneke et al. 2011). Rather than directly comparing the measured value of $a_e$ to the result of the α-dependent calculation, one can assume that the QED calculation is correct, and use the measured $a_e$ to extract a more accurate value for α:

$$\frac{1}{\alpha} = 137.035\ 999\ 1500\ (18)\ (18)\ (330) \qquad (10)$$

The test of the QED calculation, then, is a comparison of this value to the best value obtained by other means. The uncertainties in this value come from the $10^{th}$-order QED calculation, the hadronic contribution, and the experimental measurement of $a_e$, respectively.

The best independent measure of α comes from a measurement of the recoil velocity for a rubidium atom absorbing a single laser photon of momentum $\hbar k$:

$$v_{rec} = \frac{\hbar k}{m_{Rb}} \qquad (11)$$

This value can be used to find the value of α in terms of other well-known constants:

$$\alpha^2 = \frac{2R_\infty}{c}\frac{h}{m_e} = \frac{2R_\infty}{c}\frac{m_{Rb}}{m_e}\frac{h}{m_{Rb}} \qquad (12)$$

The Rydberg constant $R_\infty$ and mass ratio $\frac{m_{Rb}}{m_e}$ are known to better than ppb accuracy, so a ppb or better measurement of $v_{rec}$ becomes a precise measurement of α.

The best recoil velocity measurement (Bouchendira, 2011) to date uses Ramsey-Bordé interferometry (Bordé, 1989), which maps the phase shift of separate atomic wavepackets onto the transition probability for the atoms moving between two internal states. A sample of ultracold Rb atoms in the F=2 ground state are illuminated with a pair of lasers driving a Raman transition that performs a π/2 pulse. This leaves the atoms in an equal superposition of F=1 and F=2, but the atoms in the F=1 state have picked up a velocity of two photon recoils. A second π/2 pulse a short time later creates Ramsey fringes in the velocity distribution of the F=1 atoms; atoms remaining in F=2 are removed. A second pair of π/2 pulses some time later returns the atoms to F=2, convolving the velocity distribution with a second Ramsey fringe pattern. The probability of atoms returning to the F=2 state at the end of the Ramsey-Bordé sequence is thus sensitive to the relative phase of these two fringe patterns.

In the absence of other interactions, the phase shift detected by the Ramsey-Bordé interferometer is sensitive to accelerations or rotations of the atoms between the two sets of π/2 pulses; as a result, the technique has found application in precision sensing (Wang, 2015). The recoil velocity measurement needed for determining α is made by applying a large acceleration to the atoms between the two pairs of π/2 pulses, using a pair of counter-propagating beams with a frequency offset between them. This can be viewed as either trapping the atoms in an accelerated optical lattice, or as performing a series of *N* Raman transitions that start and end in the same internal state, but increase the atomic velocity by $2v_{rec}$ for each transition.

At the end of the acceleration stage, the atoms have acquired a velocity of $2Nv_{rec}$, which leads to a phase shift that is read out by the final pulse pair in the Ramsey-Bordé sequence. The final phase shift also includes a small contribution due to the acceleration of gravity during the acceleration stage, but this is removed by comparing the phase shifts for upward- and downward-accelerated atoms. The final measurement of the recoil velocity, combined with the Rydberg constant and Rb mass ratio gives a measurement of α with an uncertainty of 0.66 ppb (Bouchendira, 2011):

$$\frac{1}{\alpha} = 137.035\ 999\ 037(91) \tag{13}$$

This agrees with the result from the anomalous magnetic moment to within the experimental uncertainty, and confirms QED at the ppb level. At this level of precision, the QED calculation includes both muonic and hadronic contributions; omitting these terms would lead to a difference of about 2.5σ between measurements of α, constraining the possibilities for new particles and fields.

4.3 Future Prospects

Given the sheer complexity of the QED calculations, it is unlikely that searches for beyond-Standard-Model physics using the anomalous magnetic moment of the electron itself will proceed much farther in the near future. Work is underway, however, to measure the anomalous magnetic moment of the positron by the same techniques (Fogwell Hoogerheide et al., 2015); if comparable measurement precision can be obtained, this will provide a direct comparison of matter and antimatter at the ppt level, which would be a stringent test of CPT symmetry.

One of the most intriguing of several hints of possible new physics in recent years comes from the analogous measurement in another lepton, the muon. The anomalous magnetic moment of the muon is more sensitive to new physics than that of the electron, owing to the muon's greater mass, and thus a promising place to look. The muon magnetic moment was last measured experimentally at Brookhaven in the late 1990's (Bennett et al., 2006) to be:

$$a_\mu^{(ex)} = 0.001\ 165\ 920\ 80(54)(33) \qquad (14)$$

(numbers in parentheses are statistical and systematic uncertainties, respectively). The best current theoretical value (Blum et al., 2013), on the other hand, is:

$$a_\mu^{(th)} = 0.001\ 165\ 918\ 28(49) \qquad (15)$$

which is smaller than the experimental value by nearly 4σ. This discrepancy has persisted through many recalculations over the last decade.

New experiments are in preparation at Fermilab (Chapelain, 2017) and J-PARC (Kitamura et al., 2017) that should reduce the experimental uncertainty by a factor of 4. In parallel, new computational efforts aim to reduce the theoretical uncertainty by a comparable amount (Hagiwara, 2017). If both programs achieve their uncertainty goals without changing the underlying values, the discrepancy would reach the 8σ level, and represent a solid detection of new physics.

The other obvious area for improvement in precision tests of QED is the value of the fine-structure constant itself, as the uncertainty in α is one of the dominant sources of uncertainty in QED calculations of magnetic moments and other properties. Technical improvements in atom interferometry offer potential improvements in the recoil-velocity measurements (Estey et al., 2015) in the near future. An improved value of α would enable more stringent tests in a number of areas, and is thus a development to watch.

5) Time Variation of Fundamental Constants

Attempts to reconcile the Standard Model with General Relativity involve the introduction of new fields and couplings between these new fields and the fields known in the Standard Model. This has the effect of turning many of the fixed properties we associate with the Standard Model and its particles into dynamical variables, subject to change in space and time due to changes in the new physical fields (Uzan, 2011). As the transition frequencies of atoms and molecules are related to these fundamental constants, in particular the fine structure constant α and the electron-proton mass ratio µ=$m_e$/$m_p$ , sufficiently precise measurements of these transition frequencies can detect or at least constrain the variation of constants over time.

In general, the fractional time variation of an atomic transition with present-day frequency ω can be written:

$$\frac{1}{\omega}\frac{\partial \omega}{\partial t} = \kappa_\alpha \frac{1}{\alpha}\frac{\partial \alpha}{\partial t} + \kappa_\mu \frac{1}{\mu}\frac{d\mu}{dt} + \kappa_g \frac{1}{g}\frac{\partial g}{\partial t} - \frac{1}{\omega_{Cs}}\frac{\partial \omega_{Cs}}{\partial t} \qquad (16)$$

where the final term reflects the fact that all frequencies are necessarily referenced to the Cs transition frequency through the definition of the SI second, and the Cs frequency itself may change over time. The coefficients $\kappa_\alpha$, $\kappa_\mu$, and $\kappa_g$ reflect its sensitivity to changes in the fine-structure constant α, the electron-proton mass ratio µ, and the nuclear g-factor, respectively.

Each of these sensitivity coefficients is specific to a particular atomic or molecular transition; the latter two are only relevant for hyperfine transitions that depend on the magnetic moment of the nucleus. Different atomic species, and different transitions within a particular atom can have sensitivity coefficients that differ by an order of magnitude or more. The process of constraining time variation of fundamental constants, then, involves taking *ratios* of frequencies, to compare transitions that are highly sensitive to variation of the constants with transitions that are relatively insensitive.

To determine the time variation, it is obviously necessary to compare measurements at different times, to calculate an approximate value of $\frac{\partial \omega}{\partial t}$. The expected rate of change is extremely small, and the necessary sensitivity can be obtained in one of two ways: one approach uses moderately high precision measurements of values from the distant past to constrain variation over extremely long time scales, while the other uses ultra-precise laboratory measurements to constrain present-day variation on a time scale of a few years.

5.1) Changes on Cosmological Time Scales

Determining the value of fundamental constants in the distant past requires some sort of "fossil record" of physical processes as they occurred in the distant past. One of the best-known examples of this general approach comes from the "natural nuclear reactor" at Oklo in Gabon, where water seeping into uranium deposits around 1.7 billion years ago acted as a moderator, enabling fission reactions that eventually depleted the fissionable isotopes. The isotope ratios in the remaining ore reflect nuclear reaction rates at the time of the reactor's operation, and thus limit the possible variation of α over the relevant time period to $\frac{\Delta \alpha}{\alpha} \leq 1.1 \times 10^{-8}$ (Davis and Hamdan, 2015). Another record of nuclear reactions constraining time variation of α comes from the decay of 187Re into 187Os recorded in iron meteorites; this provides a weaker constraint, $\left|\frac{\Delta \alpha}{\alpha}\right| = (2.5 \pm 16) \times 10^{-7}$ (Fujii and Iwamoto, 2003).

In the atomic sector, constraints on the variation of α over time come from spectroscopic observations of objects at high redshift, either emission lines detected in the spectrum of the object itself, or absorption lines from gas and dust at an intermediate distance. These spectral features allow us to look back at transition frequencies in the distant past, to determine if they differ from the present-day values. This process is complicated by the need to account for the cosmological redshift from the Hubble expansion and any Doppler shifts due to motion of the source, so these studies generally compare multiple transition frequencies with different sensitivities to the constants of interest. Transitions with low sensitivity to changes in the fundamental constants are used to determine the redshift, which is then removed for the analysis of more sensitive transitions to determine the past value of α.

Numerous observations over the last few decades have provided measurements of possible changes in the fine structure constant, generally at the ppm level. Some of these measurements suggest that α has changed over time, others are consistent with no change. A recent review by Martins (Martins, 2017) provides an overview and some meta-analysis of the existing data.

Interest in spectroscopic constraints on changes in fundamental constants has increased significantly in recent years thanks to the dramatic suggestion by Webb et al., (Webb et al., 2011) that the discrepancies between measurements could be addressed if the value of α varies not only in time, but in space. Using observations from the Keck telescope in Hawaii and the VLT in Chile, they find a dipolar variation in α of the form:

$$\frac{\Delta \alpha}{\alpha} = Ar \cos \theta \qquad (17)$$

where $r$ is the look-back distance in Glyr, θ the angle from the pole of the dipole pattern, and $A = (1.1 \pm 0.25) \times 10^{-6} Glyr^{-1}$ (Webb et al., 2011). In this picture, the fine-structure constant was about one part per million larger one billion years ago in one direction on the sky, and one ppm smaller in the past in the opposite direction.

The claim of a definite detection of changes in α, let alone spatial variation in α, is not without controversy. Several rounds of claims and counter-claims in the literature have weakened the evidence somewhat, but not ruled out the possibility of a real dipole pattern in the value of α in the distant past, as claimed by Webb et al. (2011). A review including the most recent observations (Martins, 2017) suggests that the data are consistent with dipolar variation at about the 2σ level, which is intriguing but should be treated with caution. The astrophysical question will most likely remain open at least until the new ESPRESSO spectrograph at the VLT in Chile can provide new observations with lower uncertainties and better control of systematic errors, possibly until completion of the ELT-HIRES telescope (expected in 2024). Improved astronomical spectroscopy may also benefit from the precision techniques described in Section 3, through the use of femtosecond frequency combs as stable reference sources for calibrating spectroscopes over a wide range of frequencies in a more uniform way than is possible with traditional lamp sources (Murphy et al., 2012).

## 5.2) Present-Day Variation

Individual astronomical observations provide single values of $\Delta\alpha/\alpha$ for a particular source at some point in the past. This can be converted to an approximate rate of change $\frac{\partial \alpha}{\partial t}$, assuming a linear shift over time; a measured shift in α at the ppm level looking back $10^{10}$ years suggests a rate of change of order $10^{-16}$ yr$^{-1}$.

This inferred rate of change is on the same order as the precision of state-of-the-art fountain clocks, so monitoring the frequency ratio of two microwave clocks for a single year can in principle constrain the present-day variation of α at the same level as astronomical observations over much longer time scales. Trapped ion clocks and lattice clocks offer even greater sensitivity, with inherent precision on the order of $10^{-18}$, which approaches that needed for a test of the dipole pattern observed by Webb et al. (2011). If their observations reflect a smooth variation in the value of α continuing to the present day (as opposed to, for example, a domain-wall scenario in which the value of α is locally constant but changes abruptly at some distant boundary (Olive et al., 2011)), the motion of the solar system through that background should lead to a rate of change of order $10^{-19}$ yr$^{-1}$ and an annual modulation of order $10^{-20}$ yr$^{-1}$

The longest-running clock comparison experiment uses the cesium fountain clock at LNE-SYRTE, comparing the Cs ground-state hyperfine frequency to the analogous transition in rubidium (with a frequency of 6.8GHz) using a dual-species fountain clock. As both the Cs and Rb transitions are hyperfine transitions, the frequency ratio is sensitive to changes in both α and the electron-proton mass ratio. Analysis of some 14 years of Rb-Cs clock data (Guena et al., 2012) constrains these to $\frac{\dot{\alpha}}{\alpha} = (-2.5 \pm 2.6) \times 10^{-17} yr^{-1}$ and $\frac{\dot{\mu}}{\mu} = (15 \pm 30) \times 10^{-17} yr^{-1}$.

The best single measurement constraining time variation of α comes from a comparison between Al+ and Hg+ trapped-ion clocks (Rosenband et al., 2008). The optical-frequency transitions used in these clocks are sensitive to variations in μ and g only at negligible higher orders, and thus the ratio provides a constraint of to $\frac{\dot{\alpha}}{\alpha} = (-1.6 \pm 2.3) \times 10^{-17} yr^{-1}$ based on a single year of monitoring.

An intriguing recent development in the field is the use of two transitions in the same ytterbium ion $^{171}Yb^+$, the E2 transition at 436nm and the E3 transition at 467nm. These frequencies have been measured with reference to Cs frequency standards at PTB in Germany (Huntemann et al., 2014), and compared directly to each other by means of a frequency comb at NPL in the UK (Godun et al., 2014). When combined with earlier measurements (shown graphically in Fig. 4, from Huntemann et al. (2014)), these provide the most stringent current limit on present-day variation of the fundamental constants:

$$\frac{\dot{\alpha}}{\alpha} = (-0.7 \pm 2.1) \times 10^{-17} yr^{-1} \tag{18}$$

$$\frac{\dot{\mu}}{\mu} = (2 \pm 11) \times 10^{-17} yr^{-1} \tag{19}$$

## 5.3) Other Time-Varying Effects

A slow variation of atomic and molecular transition frequencies over time can be a signature of a universal drift in the value of fundamental constants. Many models of beyond-Standard-Model physics also introduce couplings to new fields that can lead to variations on a shorter time scale. These short-term variations can also be constrained by precision spectroscopic measurements.

One way for new physics to lead to short-term time variation would introduce a coupling between the fundamental constants and the local gravitational potential. While the eccentricity of the Earth's orbit around the Sun is small, it is large enough to provide an appreciable variation in gravitational potential, which would lead to a seasonal variation in the value of atomic transition frequencies.

The best constraints on these values come from long-term comparisons of Cs and Rb fountain clocks and hydrogen masers operating in standards laboratories around the world. The Rb-Cs comparison at LNE-SYRTE discussed above (Guena etal., 2012) check for variation of the Rb/Cs transition frequency ratio and constrains the local-position-invariance-violating coupling to $\beta = (0.11 \pm 1.04) \times 10^{-6}$, while comparisons between fountain clocks and hydrogen masers using the Cs/H and Rb/H ratios give values of $(3.6 \pm 4.8) \times 10^{-6}$ and $(6.3 \pm 10) \times 10^{-6}$, respectively (Tobar et al., 2013). Similar comparisons at USNO give limits of $(-1.6 \pm 1.3) \times 10^{-6}$ for Rb/Cs, $(-0.7 \pm 1.1) \times 10^{-6}$ for Cs/H, and $(-2.7 \pm 4.9) \times 10^{-6}$ for Rb/H (Peil et al., 2013). The most recent comparison of Cs clocks and H masers, from NIST in Boulder (Ashby et al., 2017), finds a limit of $\beta = (0.22 \pm 0.25) \times 10^{-6}$.

Data from several different measurements can be used to disaggregate the contributions of different fundamental constants, which are expressed in terms of coupling constants $k_\alpha, k_q, k_\mu$ between the gravitational potential and the fine-structure constant, electron/quark mass ratio, and electron/proton mass ratio, respectively. These are analogous but not identical to the constants in Eq. 16, as $k_\alpha, k_q$ and $k_\mu$ characterize variation not with time but with gravitational potential. The best current limits on these couplings are:

$$k_\alpha = (0.74 \pm 1.8) \times 10^{-7} \qquad (20)$$

$$k_q = (-25 \pm 21) \times 10^{-7} \qquad (21)$$

$$k_\mu = (25 \pm 54) \times 10^{-7} \tag{22}$$

where the first two values are from (Ashby et al., 2017) and the third from (Peil et al., 2013).

An independent constraint on coupling to the fine-structure constant can also be obtained from spectroscopy of atomic dysprosium, which features two nearly degenerate levels of opposite parity, thanks to an accidental cancellation of many relativistic corrections. In $^{162}$Dy, the "A" state is 236MHz above the "B" state, while in $^{164}$Dy, the two are reversed, with A below B by 754MHz. A change in the fine-structure constant would lead to a differential shift, with the level splittings of the two isotopes shifting in opposite directions. This difference allows dysprosium to be self-referencing: the differential shift between isotopes measures a change in α without the need to compare to another element, greatly simplifying the experiment and analysis.

Analysis of two years of Dy spectroscopic data gives constraints on both the time variation of the fine-structure constant, and the possible coupling to the gravitational potential (Leefer et al., 2013):

$$\frac{\dot{\alpha}}{\alpha} = (-5.8 \pm 6.9) \times 10^{-17} yr^{-1} \tag{23}$$

$$k_\alpha = (-5.5 \pm 5.2) \times 10^{-7} \tag{24}$$

(the Dy states in question are only sensitive to variations in α, not the mass ratios). This is consistent with the analysis of Ashby et al. (2017), though with a larger uncertainty.

Another new physics channel that might lead to detectable time variation of atomic and molecular transition frequencies is a weak coupling to extremely light dark matter particles (masses below $10^{-15}$ eV). If dark matter consists of bosonic particles with sub-eV masses (such as axions or dilatons), the resulting occupation numbers can be great enough to appear like a classical field, oscillating at a frequency proportional to the mass of the dark matter particle. A weak coupling between these dark-sector particles and the electromagnetic field would then manifest as an oscillation in the fine-structure constant at this same frequency. The strength of such a coupling to dark matter is constrained by existing tests of the equivalence principle, but for ultra-light masses, the sensitivity of precision AMO measurements can set new limits.

The same dysprosium system described in Leefer (2013) has been used to search for this time variation, by analyzing the power spectrum of their frequency shift signal (Van Tilburg et al., 2015). Using ten days' worth of data spread over a period of about two years, they improve on the equivalence principle constraints for a range of masses from $3 \times 10^{-18} eV$ down to $10^{-24} eV$. At their peak sensitivity, around $10^{-22}$ eV, the limit from dysprosium spectroscopy is nearly four orders of magnitude better. If an ultra-light particle in this range were to exist, its coupling to electromagnetism can be no greater than $4.2 \times 10^{-8}$ times its coupling to gravity, at the 95% confidence level.

**5.4) Future Prospects**

The best current limits from searches for present-day variation of the fundamental constants provide tight constraints on Lorentz symmetry violations and ultra-light dark matter. These experiments remain around an order of magnitude away from the sensitivity needed to detect the sort of variation that might be expected from motion through a gradient in the value of the fine-structure constant.

There are two paths to possible sensitivity improvements in searches for time-varying constants: either improving the sensitivity of the individual measurements, or increasing the time span of the search. The potential gains of the latter approach depend in large part on the amount of data already acquired. In the case of clock comparisons, a repeat of the Al+/Hg+ measurement would improve the bound on $\dot{\alpha}/\alpha$ by nearly an order of magnitude even without improvements in the spectroscopy, simply because almost a decade has passed since the original one-year measurement (Rosenband et al., 2008). The limits from Rb/Cs and H/Cs comparisons, however, are already based on 14+ years of clock signals (Ashby et al., 2017, Tobar et al., 2013), making it impractical to substantially improve these bounds simply by acquiring more data.

Improvements in the sensitivity of searches for time-varying constants are more likely to be driven by improvements in the frequency standards themselves. In recent years, optical frequency standards have been improving very rapidly, with both strontium lattices (Nicholson et al., 2015, Ushijima et al., 2015) and ytterbium ions (Huntemann et al., 2016) reaching the $10^{-18}$ level. Comparisons between different clocks at this level of precision become extremely challenging, though, as the gravitational redshift due to a 1-cm change in altitude near the surface of the Earth is of order $\frac{\delta \nu}{\nu} \approx 10^{-18}$ (Ludlow et al., 2015). This gravitational redshift issue might be evaded by comparing transitions within a single ion as in Yb+ (Godun et al., 2014), or using ions of two different species held in the same trap.

An intriguing possibility for a new system to test the stability of fundamental constants is a "nuclear clock" based on an isomer transition in thorium-229, a proposal that has been extensively discussed in recent years. The nuclear energy levels making up the clock in this system provide excellent isolation from most environmental effects that lead to systematic shifts in electronic transitions, and the transition wavelength should fall in the vacuum ultraviolet range, making it accessible with existing laser technology. Estimates of the clock performance suggest that it could operate at the $10^{-19}$ level of accuracy.

While the isomer state has been been confirmed through detection of the electrons produced in internal conversion decays on a surface (von der Wense et al., 2016), it has not yet been observed spectroscopically, despite many years of searching (Yamaguchi et al., 2015). There is also significant uncertainty regarding the sensitivity of this transition to changes in α, with some estimates suggesting a large sensitivity enhancement (Flambaum, 2006), and others basically none (Hayes et a., 2008). Whether a thorium-ion clock can be useful in the search for new physics thus remains an open question.

A final system attracting interest for precision metrology applications, including new-physics searches, is spectroscopy of highly charged ions. In general, atomic transition frequencies increase rapidly with the degree of ionization, pushing most electronic transitions into the x-ray region of the spectrum, beyond the reach of current laser technology. In some highly ionized systems, though, new laser-accessible transitions can appear.

For very heavy elements, the removal of electrons can cause the re-ordering of electronic states that are ordinarily shifted by intra-atomic interactions. For example, in Ag-like ions, the 4f state moves below the 5s state in energy for atomic numbers greater than about Z=61 (Berengut et al., 2013). In the vicinity of such a level crossing, there can be new optical-frequency transitions in highly charged ions with ionization energies running to hundreds of eV.

At extreme levels of ionization, the hyperfine splittings of hydrogen-like ions can be increased, bringing them into the range of infrared lasers (Schiller, 2007). These systems would be sensitive to variations in the mass ratio µ as well as α. They can also enable other types of new-physics searches, such as tests of strong-field QED. Laser spectroscopy measurements of the hyperfine splitting in $Bi^{82+}$ and $Bi^{80+}$ show a large discrepancy with theoretical predictions (Ullmann et al., 2017), though further measurements are needed to rule out experimental systematics as the source of this hyperfine puzzle.

Several highly-charged ion systems have been identified as having attractive properties for precision metrology applications (Derevianko et al., 2012; Yu and Sahoo, 2016; Nandy and Sahoo 2016), and in particular for searches for time-varying constants (Ong et al, 2014). The level crossings of interest have been studied using ions in electron beam ion traps (Windberger et al., 2015), in particular the $Ir^{17+}$ system that has been calculated to be exceptionally sensitive to changes in α. Sympathetic cooling of $Ar^{13+}$ by $Be^+$ ions has recently been demonstrated (Schmöger et al, 2015), opening the door for high-precision spectroscopy of cold trapped ions.

## 6) Electric Dipole Moment (EDM) Searches

The final class of precision-measurement based searches for new physics that we will discuss is the search for a permanent electric dipole moment (EDM) of a particle or nucleus. In addition to revealing the presence of beyond-Standard-Model particles and fields, such a measurement is directly related to the strong CP problem and explaining the observed overabundance of matter vs. antimatter. A permanent EDM for a particle or nucleus would necessarily violate time-reversal symmetry, as first noted by Purcell and Ramsey (1950), as the spin magnetic moment is odd under time reversal while an EDM would be even. The existence of such a T-violating EDM would thus require CP violation elsewhere in order to preserve CPT symmetry overall.

Known sources of CP violation in the Standard Model would allow an extremely small EDM for fundamental particles. In the case of the electron, the Standard Model prediction is of order $d_e \sim 10^{-40} e \cdot cm$, far too small to detect experimentally. Extensions to the Standard Model almost inevitably introduce new CP-violating phases leading to an electron EDM that is many orders of magnitude larger, of order $d_e \sim 10^{-25} - 10^{-30} e \cdot cm$, making an electron EDM a promising target for experimental searches (Engel et al., 2013).

## 6.1) Experimental Technique

The effect of a permanent EDM for a fundamental particle such as an electron is to add an orientation dependence to its energy in an external electric field $\vec{E}$ leading to an energy shift $\Delta E_{edm} \approx \vec{d}_e \cdot \vec{E}_{applied}$. For a charged particle, any EDM shift would be tiny compared to the Coulomb interaction, and the particle will simply move in response to the external field. EDM searches thus look for small energy shifts in the energy of bound states of atoms and molecules.

The key to detecting such an energy shift is the ability to apply a large electric field to the electron, a process which introduces some experimental complications. In the simplest approximation, the EDM shift in an atom or molecule also ought to be zero, as the electron orbitals should shift to cancel the applied field within the atom. For extremely heavy elements, though, relativistic effects prevent the complete cancellation of the external field, and can actually lead to an enhancement of the effective field experienced by an electron. For the best atom-based EDM measurement, in thallium, the effective field was ~580 times greater than the laboratory field (Regan et al., 2002).

The experimental situation is even more favorable within polar molecules, which can feature very strong internal electric field. A modest laboratory field (of order 100 V/cm) can effectively polarize the molecules completely, leading to an effective field for electrons within the molecule approaching 100GV/cm. For this reason, current state-of-the-art EDM searches predominantly use polar molecules.

Symmetry considerations require that any electron EDM be along the spin axis of the electron, so the generic EDM search, illustrated in Fig. 5, is a measurement of a precession frequency. Molecules are placed in a combination of (anti)parallel electric and magnetic fields, and then excited to a superposition of $m = \pm 1$ states. The relative phase of these states evolves at a frequency that depends primarily on the Zeeman effect from the magnetic field, increased or decreased slightly by the EDM shift due to the electric field. In the Bloch sphere picture, the magnetic moment of the electron is rotated to the equator, and then precesses about the applied electric and magnetic fields at a frequency proportional to the energy shift.

The combined energy shift is measured after some free evolution time either by completing a Ramsey-type pulse sequence and measuring the probability of returning to the original ground state (Regan et al., 2002; Hudson et al., 2011, Cairncross et al., 2017), or by directly measuring the spin orientation (Baron, et al., 2014; Graner et al., 2016). To separate the EDM shift from the Zeeman effect, and exclude various systematic effects, the experiment is then repeated with the relative directions of $\vec{E}$ and $\vec{B}$ reversed, reversing the perturbation due to the EDM. The difference between the frequencies for parallel and anti-parallel $\vec{E}$ and $\vec{B}$ is then twice the EDM shift.

**6.2) Experimental Results**

The best atom-based limit on the electron EDM (Regan, 2002) used $^{205}$Tl atoms with an applied electric field of $1.2 \times 10^5 V/cm$. To reject systematic effects relating to motional magnetic fields, the experiment used two pairs of counter-propagating atomic beams, with each pair exposed to the same magnetic field but opposite electric field. To avoid issues from stray magnetic fields, the atomic beams contained a mix of thallium and sodium as a "co-magnetometer"; the low mass of Na means that it is not sensitive to an electron edm, but does experience a Zeeman shift. The final bound on the electron edm is $|d_e| \leq 1.6 \times 10^{-27} e \cdot cm$ (at the 90% confidence level).

The best current measurement comes from the ACME collaboration, using thorium monoxide (ThO) molecules (Baron et al., 2014; Baron et al., 2017). The $H^3\Delta_1$ electronic state of ThO offers an exceptionally large internal electric field, $|E_{eff}| \approx 80 \ GV/cm$, and is polarized by a small applied field in the laboratory, of order 100 V/cm, avoiding many of the systematic effects associated with large laboratory fields. The H state also offers a great advantage in having an "omega doublet" level structure with nearly degenerate levels of opposite parity, corresponding to states with the molecular axis aligned parallel or anti-parallel to the electric field axis, as shown in Fig. 6. These different orientation states shift in opposite directions in response to the applied electric and magnetic fields, providing an internal co-magnetometer, and allowing effective "reversals" of the field by switching between orientation states.

ThO molecules from a cryogenic source are excited to the *H* state in a particular molecular alignment, and prepared in a superposition of the $M = \pm 1$ states by dark-state optical pumping. After a free precession time of about 1ms, the spin orientation is measured by polarization-sensitive excitation to the molecule's C state, which decays back to the ground state by emitting a 690nm fluorescence photon that provides the signal for the experiment. Based on about 2 weeks of data, the upper bound for the electron EDM from ThO spectroscopy is $|d_e| \leq 9.4 \times 10^{-29} e \cdot cm$ (Baron et al., 2017).

The most recent limit comes from molecular ions of $^{180}$Hf$^{19}$F$^+$ held in a radio-frequency ion trap (Cairncross et al., 2017). As with trapped-ion frequency standards discussed above, the confinement of these molecules offers the benefit of much longer interrogation times, over 700ms of spin precession. Of order 1000 molecules in the trap are subjected to a rotating electric field, and all excitation and detection operations are synchronized with this rotating field. A quadrupolar magnetic field gradient added to the trap gives an applied magnetic field for the molecules that is either parallel or antiparallel to the electric field.

Trapped ions are excited to the $^3\Delta_1$ $F = \frac{3}{2}$ state, which like the H state in ThO consists of a closely spaced doublet corresponding to two different alignments of the molecular axis relative to the rotating field. The molecules are prepared in a superposition of the $m_F = \pm \frac{3}{2}$ states using a π/2 pulse applied through a rotation-induced coupling in the trap, and allowed to precess for several hundred ms before a second π/2 pulse completes a Ramsey interferometry sequence mapping the superposition back to the population difference between $m_F$ states. The population difference is measured by state-selective photoionization of the molecules, providing the experimental signal.

Based on ~300 hours of collected data, the HfF+ experiment provides a 90% confidence limit upper bound of $|d_e| \leq 1.3 \times 10^{28} e \cdot cm$ (Cairncross et al., 2017), comparable to that from the neutral ThO system. Given the radically different systematics of the trapped-ion measurement, this offers a valuable independent confirmation of the earlier limit. The HfF measurement is primarily limited by statistics, largely due to the relatively small number of trapped ions.

The final example of an EDM search constraining new physics is the University of Washington's project on a nuclear EDM of $^{199}$Hg (Graner et al., 2016). This experiment uses mercury vapor in paraffin-coated cells, with spin precession times of nearly 200s. The nuclear spin is polarized by optical pumping with a laser at 254nm, and the spin direction is determined by measuring the Faraday rotation of a weak probe laser passed through the cell. Based on some 250 days of data, the 95% confidence limit for the nuclear dipole moment is $|d_{Hg}| < 7.4 \times 10^{-30} e \cdot cm$.

As in the case of searches for time-varying constants discussed above, the involvement of nucleons adds a number of additional new physics channels that are constrained by the limit on $d_{Hg}$. The result can be interpreted as a limit on the EDM of the neutron, proton, or up and down quarks, or various CP-violating interactions between nucleons or between nucleons and electrons. If interpreted as a limit on the neutron EDM, they find $|d_n| \leq 1.6 \times 10^{-26} e \cdot cm$, a factor of two better than previous measurements on free neutrons (Pendlebury et al., 2015).

**6.3) Future Prospects**

Converting a limit on the electric dipole moment of a particle or nucleus to a limit on the properties of beyond-Standard-Model particles is a model-dependent process, but the current experimental limits place significant constraints on the possible properties of CP-violating new particles. For the most straightforward supersymmetric extensions of the Standard Model, in which the new particle interactions leading to an EDM enter at one-loop order, the ThO and $^{199}$Hg measurements would require these particles to have masses of order 10TeV. More complicated theoretical scenarios that push most supersymmetric partners to higher mass, keeping only a few light particles, weaken this constraint to a lower bound of order ~2TeV. The mass bound implied for multi-Higgs models is of a similar scale (Safronova et al., 2017). All of these constraints exceed the immediate reach of direct creation experiments at the LHC.

All of the recent EDM search experiments have clear paths forward to improve their statistical sensitivity. The ACME collaboration is implementing higher efficiency state preparation and detection systems (Panda et al., 2016), and the Imperial College YbF experiment is similarly upgrading their molecule source and detection systems (Rabey et al., 2016). The HfF+ experiment is increasing the size of their ion trap to hold more molecules, and thus improve their statistics. The 199Hg experiment is the longest-running of the major experiments, and thus has the least room for dramatic improvements, but technical upgrades to increase sensitivity by a factor of 2-3 should still be possible.

In the next 3-5 years, it is reasonable to expect the sensitivity of electron EDM searches to improve by at least an order of magnitude. This will either lead to an affirmative detection of a non-zero EDM, or exclude essentially all currently popular supersymmetric extensions of the Standard Model at a level well beyond the reach of the LHC. In the somewhat longer term, numerous experiments are proposed or in development to use new experimental configurations (Vutha et al., 2010; Tarbutt et al., 2013), or new atomic or molecular species (Kozyrev and Hutzler, 2017; Cairncross et al., 2017; Weiss et al., 2003, Inoue et al., 2015, Bishof et al., 2016).

**7) Conclusion**

The above discussion covers only a subset of the current and proposed experimental searches for new physics using high-precision atomic and molecular spectroscopy. A more comprehensive review, covering many categories of experiment not discussed here, may be found in Safronova et al. (2017).

While some of these precision-measurement searches have histories stretching back 30 years or more, many current experiments originate in the last 15 years. Interest in this subfield has expanded very rapidly over this time thanks to a fortuitous confluence of two larger trends, one positive and one negative. On the positive side, there has been steady improvement in precision metrology enabled by developments in laser technology, computational power, and experimental techniques for manipulating atoms and molecules. At the same time, beyond-Standard-Model physics has continued to elude the reach of direct-detection experiments, even as evidence has mounted pointing to the existence of a large "dark sector" of physics not explained by the Standard Model.

The increasing sensitivity of atomic and molecular spectroscopy and the increasing lower limit for new-particle masses combine to make this an extremely interesting time for precision-measurement searches for new physics. If supersymmetric partner particles had masses in the originally expected ranges, they would've been detected in accelerators long before the experimental sensitivity advanced to the point of being able to infer their presence from spectroscopic measurements. The present situation has created a window, for the moment at least, where improved metrology allows searches that might reveal new particles and fields in energy ranges that have not yet been experimentally excluded (and, in the case of some EDM experiments, ranges that may remain beyond the reach of accelerator-based searches for many years to come).

The next 3-5 years should be a particularly fertile time for this subfield of AMO physics, as essentially all of the major new-physics search experiments have clear paths forward to increase their sensitivity, in some cases by multiple orders of magnitude. If the imminent upgrades produce a definitive detection of new physics, the precision measurement community will be well positioned to study the essential phenomena from a wide range of perspectives, providing a wealth of information. If, however, the generation of experiments now in progress and in development do not find new physics, explaining that null result will surely tax the ingenuity of particle theorists.

**Acknowledgements:** Thanks to Sabine Hossenfelder for pointers on quantum gravity, Eric Copenhaver for information on the status of photon recoil measurements, and Tom Swanson and Dave Phillips for helpful discussions of precision measurements generally.

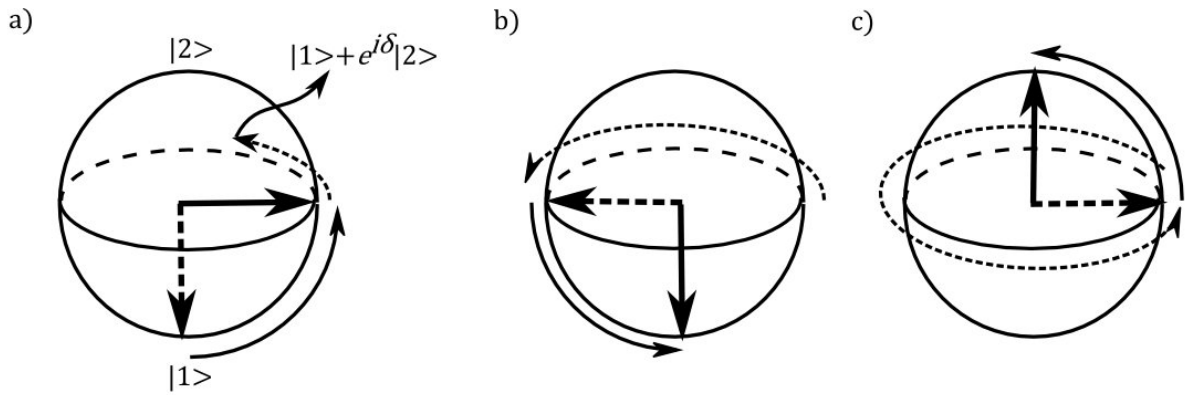

**Figure 1:** Bloch sphere visualization of the Ramsey separated-fields method. a) The initial $\pi/2$ pulse rotates the state vector into a superposition state, which begins to precess about the axis at a frequency $(\omega-\omega_0)$. b) After a half-integer number of rotations, the second $\pi/2$ pulse rotates the state vector back to state $|1\rangle$. c) After an integer number of rotations, the second $\pi/2$ pulse completes the transition to state $|2\rangle$

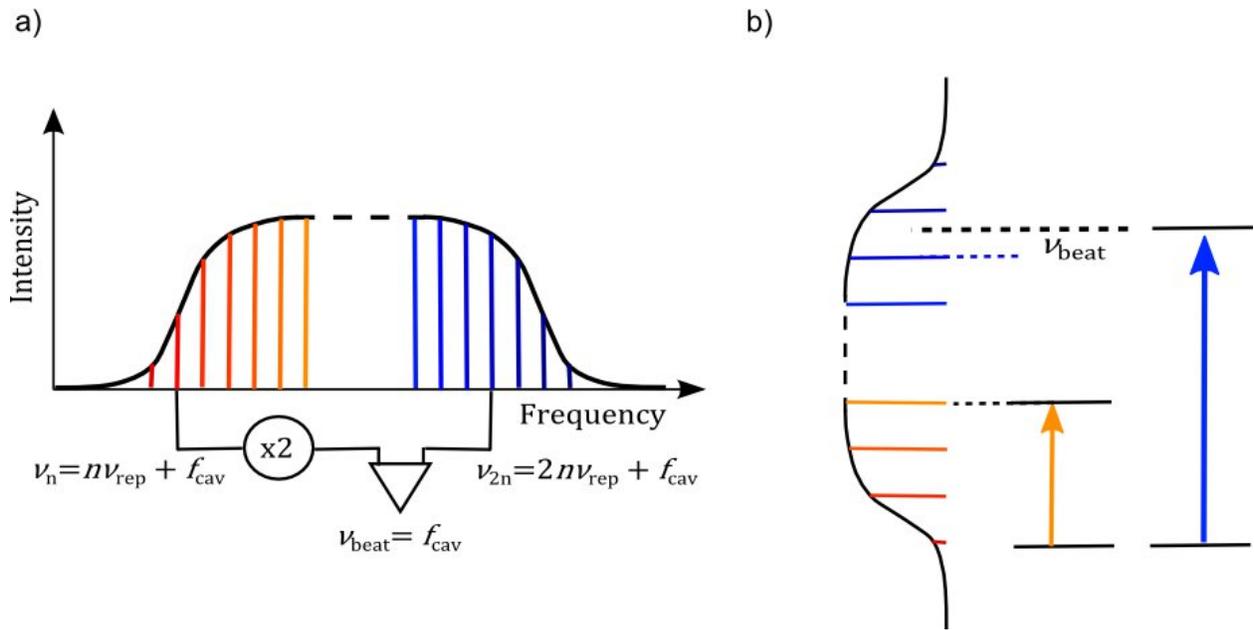

**Figure 2:** a) Self-referencing of a femtosecond frequency comb. The beat note between a high-frequency comb mode and a frequency-doubled mode from an octave lower is equal to the offset frequency due to cavity dispersion, and allows absolute determination of the frequency of any mode. b) Frequency ratio measurements using a comb. The comb is stabilized with reference to one atomic transition frequency, and the beat note between a laser locked to another atom and the nearest comb mode determines the frequency of the second laser.

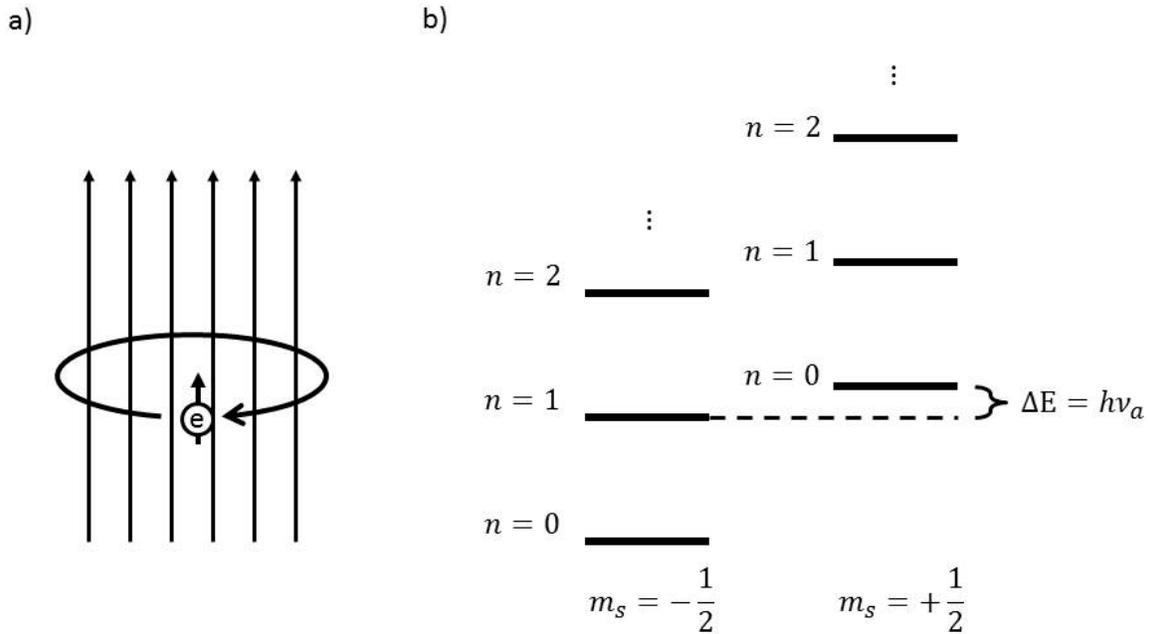

**Figure 3:** Measurement of the anomalous magnetic moment of the electron. a) Schematic of the electron in the trap, which orbits the applied magnetic field at the cyclotron frequency, and experiences a magnetic shift due to the spin alignment. b) Energy levels of the trapped electron with regularly spaced cyclotron orbits in each of the two spin states. The offset between states that differ by both a cyclotron transition and a spin flip is determined by the anomalous magnetic moment.

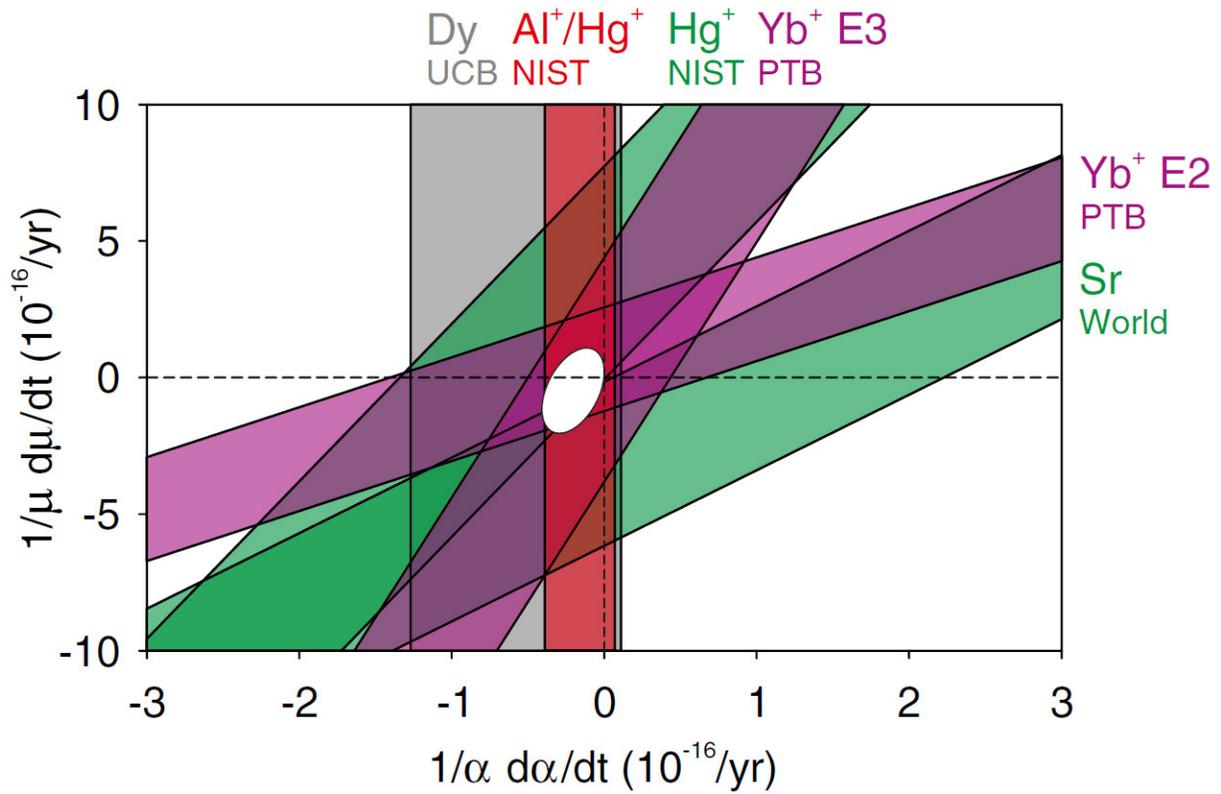

**Figure 4:** Constraints on temporal variations of α and μ from comparisons of atomic transition frequencies. Filled stripes mark the 1σ-uncertainty regions of individual measurements and the central blank region is bounded by the standard uncertainty ellipse resulting from the combination of all data. Figure from Huntemann et al., 2014 "Improved Limit on a Temporal Variation of mp/me from Comparisons of Yb+ and Cs Atomic Clocks," Phys. Rev. Lett. 113, 210802, used with permission.

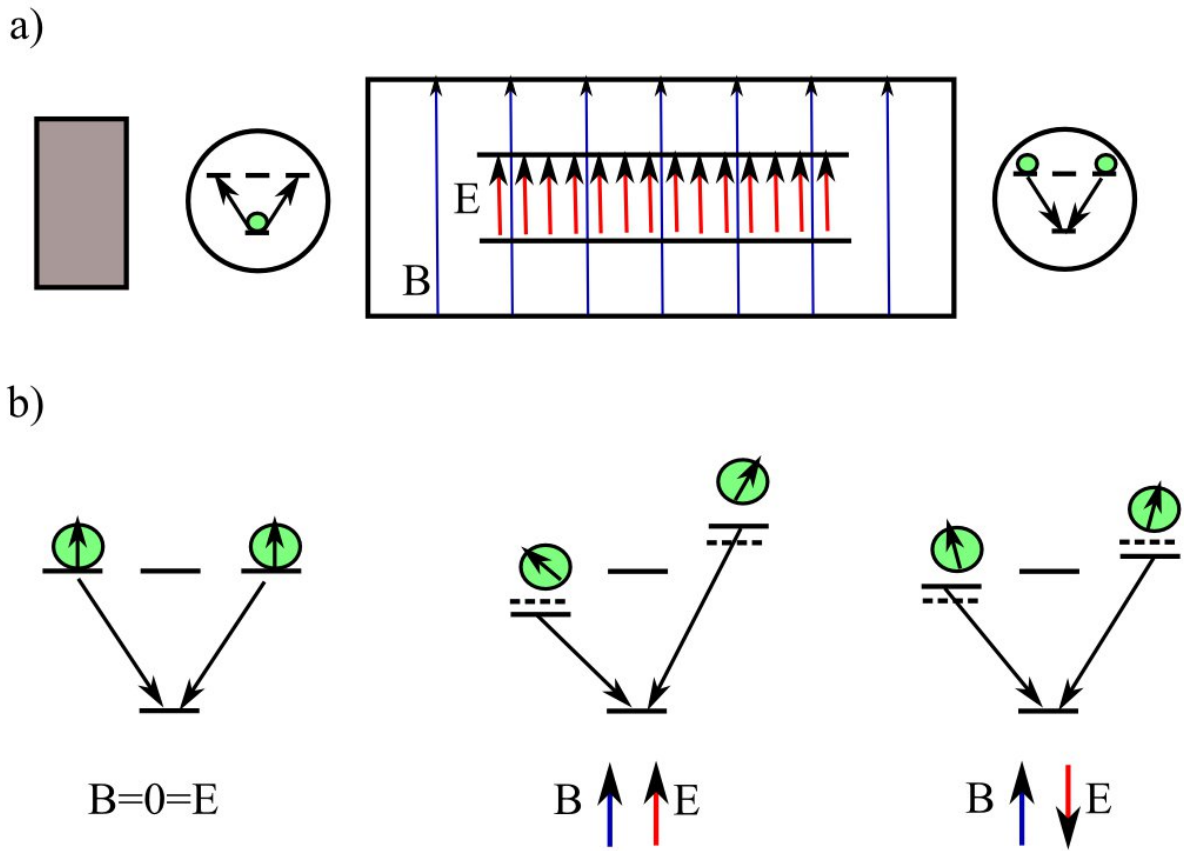

**Figure 5:** a) Experimental schematic of an electron EDM search. Atoms or molecules are prepared in an initial superposition state, then evolve freely in a region of (anti)parallel electric and magnetic fields. After the free evolution, the phase difference between the two component states is projected onto a population difference that is measured. b) Schematic of the phase measurement. With parallel E and B fields (center), the EDM shift adds to the Zeeman shift (dashed line) to produce a larger phase difference. With antiparallel E and B, the EDM shift reduces the phase difference. The change in phase on reversal of the fields is the signature of an EDM.

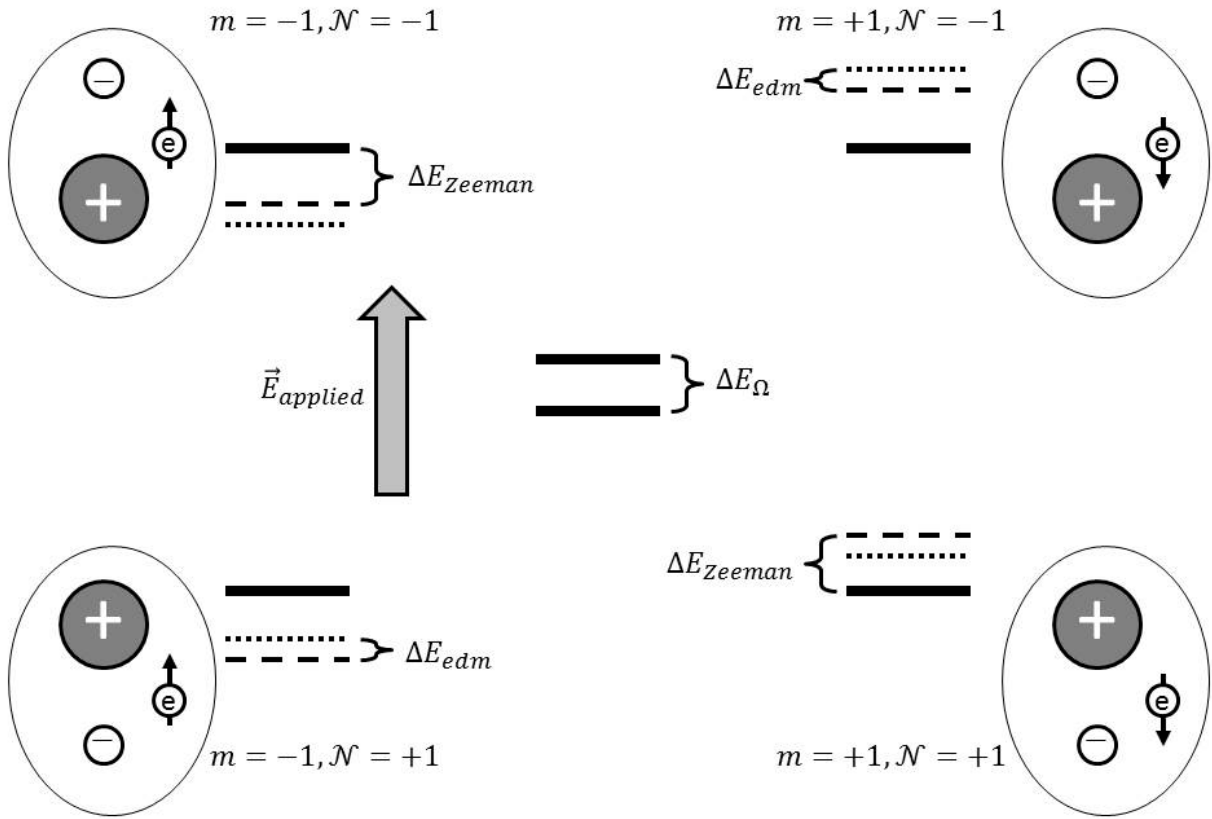

**Figure 6:** Omega-doublet level scheme for the $^3\Delta_1$ states used in the ThO and HfF$^+$ EDM searches, featuring nearly degenerate levels that differ in the orientation of the molecular axis relative to the applied fields. The Zeeman and EDM energy shifts are in different directions for different magnetic sublevels, allowing effective reversal of the field orientation by changing internal states.